# Multiple quasi-phase-matched second-harmonic generation in phase reversal optical superlattice structure

Toijam Sunder Meetei[1] · Meerasha Mubarak Ali[1] · Shanmugam Boomadevi[2] · Krishnamoorthy Pandiyan[1]

## Abstract
Domain-engineered quasi-phase-matching (QPM) devices are known for its versatility and ability to tune the nonlinear optical frequency conversion process. In this paper, a simple approach is presented to generate multiple quasi-phase-matched second-harmonic generation (SHG) in the phase reversal optical superlattice (PROS) structure. Theoretical studies are carried out by simulation based on the domain reversed lithium niobate QPM devices. The nature of the generated multiple SHG spectra is analyzed when the phase reversal (PR) domains are distributed at equal and unequal intervals along the length of the device. The distribution of phase reversal domains at equal intervals is limited to the generation of dual SHG peaks irrespective of its number. On the contrary, we could generate equal-intensity multiple SHG peaks with PR domains distribution at unequal interval. Using this scheme, five peaks QPM SHG are generated by distributing four PR domains in specific locations of the PROS QPM device. The dependency of the PR domain and its location in the PROS QPM device are analyzed to design desirable multi-wavelength converters.

## 1 Introduction

Second-harmonic generation (SHG) is a well-known and extensively studied second-order ($\chi^2$) nonlinear optical process that has enabled to access a wide range of applications in the optical frequency conversion such as imaging microscopy, wavelength division multiplexing (WDM) networks, visible and mid-IR generations [1–9]. Highly efficient all-optical wavelength converter is the highly demanded tool in an optical communication system. It is due to the flexibility and capability to deliver the data payload simultaneously from the source to the destination without being converted from an electronic domain in the optical fiber-based WDM network [10]. In this context, quasi-phase matching (QPM) devices play a vital role. The phase matching capability of the conventional periodic QPM devices is restricted to a single-phase matching condition due to the limited availability of the reciprocal vector [11]. Hence, other structures have been proposed to phase-match many wavelengths simultaneously in a single device. Otherwise, an array of wavelength converters would require for many wavelengths, which is expensive and voluminous. Therefore, multiple wavelength conversion in a single device offers more advantage in the cost reduction and miniaturization that can be easily integrated into the optical communication system or a photonic circuit.

Various domain-engineered structures have been proposed with the aim of achieving efficient multiple QPM devices, which includes aperiodic optical superlattice (AOS) structure, Fibonacci optical superlattices (FOS) and a continuously phase-modulated (CPM) grating structure [12–15]. Chou et al. [16] has successfully illustrated the multiple-channel wavelength conversion of 1.5 μm, and 1.3–1.5 μm in 42-mm and 36-mm-long domain-engineered LiNbO$_3$ waveguides respectively for the application in WDM systems. Using CPM grating structure, Asobe et al. [17] has also demonstrated a flexible way of generating multiple QPM by appropriate phase modulation in the device that achieved fast (< 100 ps) 4-channel 40 Gb/s signal wavelength switching. Most importantly, these structures are more complicated to fabricate and prone to duty cycle variations as the device possesses different grating periods along the channel

✉ Krishnamoorthy Pandiyan
krishpandiyan@ece.sastra.edu

[1] Present Address: Centre for Nonlinear Science and Engineering (CeNSE), School of Electrical and Electronics Engineering, SASTRA Deemed to be University, Thanjavur, Tamil Nadu 613401, India

[2] Department of Physics, National Institute of Technology, Tiruchirappalli, Tamil Nadu 620005, India

[18–20]. Hence, simple device structures are favorable to subside the amount of possible errors during the fabrication.

To address this issue, we present a simple QPM structure which can achieve efficient multiple wavelength conversions than the aforesaid structures. In this proposed phase reversal optical superlattice (PROS) structure, a limited number of phase reversal (PR) or aperiodic domains of length equal to the grating period ($\Lambda$) or twice the coherence length ($2l_c$) of the device is distributed in specific locations without disturbing the overall period of the device. The schematic of the PROS device is shown in Fig. 1. However, proper distribution of PR domains in the device structure is mandatory to obtain the desired multiple peaks. Using Fibonacci logic [21, 22] and the simulated annealing algorithm [23], PR domains were distributed along the length of the PROS and achieved multiple QPM successfully. However, in both cases, the major limitation is uneven multiple peak nature. Hence, we analyzed an approach of domain engineering by distributing PR domains at the specific positions in the device structure to achieve and design desirable effective multiple QPM wavelength converters. Though it is a tedious process, it greatly improves the tunability of the device.

## 2 Theory

Optical radiation of an operating or a fundamental wavelength ($\lambda_\omega$) propagating in the QPM device results in the generation of higher harmonics due to the nonlinear interaction between the interacting waves in the medium. The expression for the amplitude of the generated second harmonic (SH) can be derived by applying Fourier transform to one of the coupled modal equations under negligible fundamental depletion condition [11]. The fundamental form of slowly varying amplitude equation governing for the growth of SH in the nonlinear crystal which represents the frequency up-conversion process is given as

$$\frac{dA_2}{dz} = i\kappa A_1^2 e^{-i\Delta k z} \tag{1}$$

In Eq. 1, $A_1$ and $A_2$ are the fundamental and second-harmonic electric field amplitudes propagated in the medium or crystal, respectively. Also, $\kappa = \frac{\omega d(z)}{c n_2}$ is proportional to the effective nonlinear optical coefficient ($d_{eff}$). To obtain effective nonlinear optical conversion, the phase of all the interacting waves traveling in the crystal must be maintained (i.e., same phase velocity) along the direction of the propagation of waves [24]. However, there exists a varying phase relationship due to the intrinsic dispersive property in the materials [25], and the arising phase-mismatch is given as,

$$\Delta k = k_2 - 2k_1 \pm G, \tag{2}$$

where, $G$ is the grating vector, $k_1$ and $k_2$ are the wavenumbers of the fundamental and SH waves, respectively. This phase-mismatch can be compensated by implementing the QPM technique. Further, with the assumption of negligible depletion of the fundamental wave for the convenience of our theoretical calculation, Eq. 1 can be integrated to obtain a *sinc*-form of the spectrum provided $\kappa$ is assumed as constant in the medium. However, $\kappa$ is not a constant and it is a function of $z$ i.e., $\kappa(z)$. Consider $\Delta k = 2\pi q$, the SH amplitude can be rewritten as the Fourier transform of $\kappa(z)$,

$$A_2(q) = iA_1^2 \int \kappa(z) e^{-i2\pi q z} dz = iA_1^2 \mathfrak{I}[\kappa(z)] \tag{3}$$

Further, the phase reversal (PR) domain(s) $w_{i=1,\dots,N}$ of width $\Lambda$ is introduced in the conventional periodic QPM structure. This distribution of one aperiodic/PR domain at the location $L/2$ (center of the device) in the periodic structure of the device gives rise to the generation of dual SH peaks near the phase matching wavelength [26]. Therefore, the effective nonlinear coefficient term $\kappa(z)$ can be represented as

$$\kappa(z) = d_{\text{eff}} \sin\left(\frac{2\pi z}{\Lambda}\right) \left[\text{rect}\left(\frac{z - \frac{L}{4}}{\frac{L}{2}}\right) - \text{rect}\left(\frac{z - \frac{3L}{4}}{\frac{L}{2}}\right)\right] \quad (4)$$

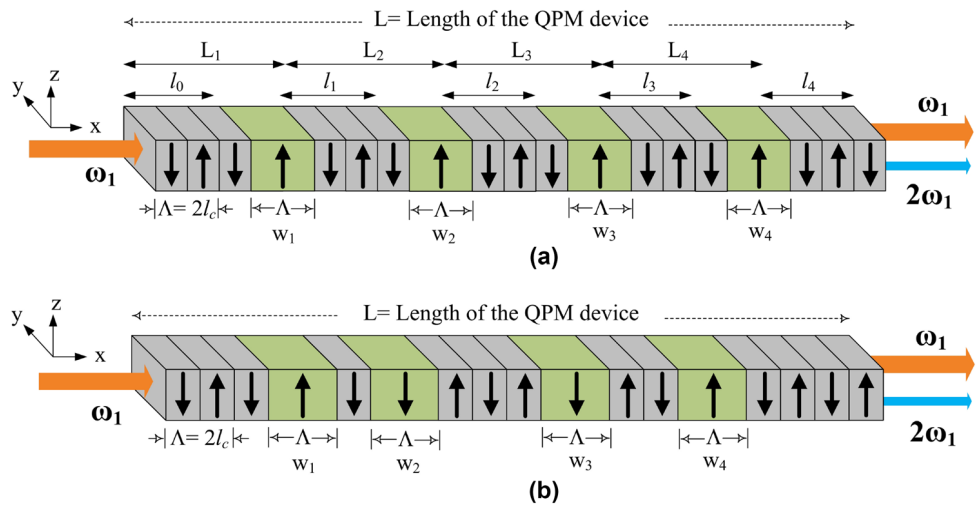

**Fig. 1** Pictorial representation of the PROS QPM device where the PR domains are distributed **a** at an equal interval and **b** unequal interval along the device length

Now, the Fourier transform of $\kappa(z)$ in Eq. 4 can be written as

$$\Im[\kappa(z)] = \frac{i}{2} L d_{\text{eff}}\left(q \pm \frac{1}{\Lambda}\right) \left[\text{sinc}\left(\frac{Lq}{2}\right) e^{i\frac{\pi}{2}Lq} - \text{sinc}\left(\frac{Lq}{2}\right) e^{i\frac{3\pi}{2}Lq}\right] \quad (5)$$

Substituting Eq. 5 in Eq. 3, the SH amplitude ($A_2$) of the QPM device with PR domains can be rewritten as,

$$A_2(q) = iaLA_1^2 e^{i\pi L\left(q \pm \frac{1}{\Lambda}\right)} \frac{\sin^2\left\{\frac{\pi}{2}L\left(q \pm \frac{1}{\Lambda}\right)\right\}}{\frac{\pi}{2}L\left(q \pm \frac{1}{\Lambda}\right)} \quad (6)$$

Equation 6 can be further reduced to,

$$A_2(q) = iaLA_1^2 e^{ip\chi'} \frac{\sin \chi' \sin p\chi'}{\chi' \cos \chi'} \quad (7)$$

In Eq. 7, $\chi' = \pi l(q \pm 1/\Lambda)$, $p = L/m\Lambda$, $L$ is the length of the QPM device, $m$ is the total number of domains, and $\Lambda$ is the grating period. The distribution of one aperiodic domain at the location $L/2$ in the periodic structure of the device was limited to the generation of dual SH peaks. Therefore, we implemented an approach of random distribution of N PR domains at suitable locations along the length of the device to overcome and generate desired multiple SH peaks. The effective nonlinear coefficient $\kappa(z)$ of a QPM device with N PR domains of a PROS structure is given by,

$$\kappa(z) = d_{\text{eff}}\left[\text{rect}\left(\frac{z - \left(\frac{l_o}{2}\right)}{l_o}\right) \sin\left(\frac{2\pi}{\Lambda}\right) + \sum_{i=1}^{N}\left\{(-1)^i\left\{\text{rect}\left(\frac{z - \left(L_i + \frac{w_i}{2}\right)}{l_i}\right)\right.\right.\right.$$
$$\left.\left.\left.\sin\left(\frac{\pi}{w_i}\right) + \text{rect}\left(\frac{z - \left(L_i + \frac{w_i}{2}\right)}{l_i}\right) \sin\left(\frac{2\pi}{\Lambda}\right)\right\}\right\}\right] \quad (8)$$

In Eq. 8, $l_o$ is the length of the first periodic region in the device, $l_{i=1…N}$ is the length of the remaining periodic regions in the device, $L_{i=1…N}$ is the length of the segment as shown in Fig. 1. Here, each segment is separated by a PR domain. Then, the Fourier transform of $\kappa(z)$ in Eq. 8 with N PR domains introduced in the periodic QPM structure is defined as

$$\Im[\kappa(z)] = \frac{1}{2} d_{\text{eff}}\left[l_o \text{sinc}\left\{l_o\left(q \pm \frac{1}{\Lambda}\right)\right\} e^{i2\pi \frac{l_o}{2}\left(q \pm \frac{1}{\Lambda}\right)}\right.$$
$$+ \sum_{i=1}^{N}\left\{(-1)^i\left\{w_i \text{sinc}\left\{w_i\left(q \pm \frac{1}{2w_i}\right)\right\} e^{i2\pi\left(L_i + \frac{w_i}{2}\right)\left(q \pm \frac{1}{2w_i}\right)}\right.\right.$$
$$\left.\left.\left. + l_i \text{sinc}\left\{l_i\left(q \pm \frac{1}{\Lambda}\right) e^{i2\pi(L_i \pm \frac{l_i}{2})(q \pm \frac{1}{\Lambda})}\right\}\right\}\right] \quad (9)$$

Distribution of PR domains in specific locations is very important to obtain a symmetric multiple SHG spectral response. Two schemes to distribute PR domains have been proposed in this work, i.e., distribution of equal and unequal intervals as illustrated in Fig. 1a, b, respectively. In the first scheme of PR domains distributed at an equal interval, for N number of PR domains, the location of PR domain in the device is given as $m/(N + 1)$, where $m$ is the number of domains present in the device. For example, when $N = 3$ is distributed in 5-mm-long QPM device of 20 µm grating period ($\Lambda$), the number of domains present in the device is 500 of each domain length 10 µm ($l_c = \Lambda/2$). Therefore, the location of the three PR domains is 125, 250 and 375. The segment length ($l_i$) between the consecutive PR domains is equal, i.e., 500/4 = 125. In the second scheme, the suitable location of the PR domains is acquired by slightly changing the positions of the PR domains obtained in the first scheme. Therefore, the distance between the consecutive PR domains distributed in the device is different. For our simulation, we considered a type-0 (e + e → e) one-dimensional QPM SHG interaction that accesses the highest nonlinear coefficient $d_{33}$ in congruent lithium niobate (CLN) crystal. This type-0 interaction is almost insensitive to the temperature change. The refractive indices of the material that depend on the wavelength of the light were calculated by Sellmeier equation [27]. The domain walls are considered to be straight in nature throughout the analysis to reduce the complexity in our simulation.

## 3 Results and discussion

For the entire analysis, we considered a 10-mm-long CLN-based PROS QPM device of 20 µm grating period. A total of 1000 domains of each domain length 10 µm are considered in the structure. First, the SH spectral response was analyzed for the PROS QPM devices by distributing PR domains in the periodic structure at equal intervals. In this case, the distance between the consecutive PR domains are equal as shown in Fig. 1a. The PR domains, i.e., $N = 1, 2, 3$ and $4$ of the same width ($w = 20$ µm) are distributed at equal intervals in the QPM device. The normalized SH intensity responses are shown in Fig. 2a–d. The inclusion of PR domains in the periodic structure resulted in equal-intensity dual peaks near the phase matching point due to the availability of reciprocal vector and phase variation [28]. The interleaved peaks appeared in the SHG spectrum shows the indication of additional peaks possible for a specific number of PR domains. For example, in Fig. 2d, when $N = 4$, PR domains are located at 200th, 400th, 600th and 800th positions in the device, i.e., the distance between the location of consecutive PR domain is constant which is also depicted in Fig. 3a. In this case, a total of five peaks are observed with two prominent equal intensity SH peak at fundamental wavelengths 1.554 µm and 1.624 µm and three interleaved peaks at 1.573 µm, 1.588 µm and 1.603 µm. Hence, taking into account of all the interleaved peaks and the dual peaks from the SHG spectra, we inferred that $N + 1$ SH peaks are generated by following the first proposed scheme of PR domains distribution at equal interval. The side lobes/bands are ignored in our studies as its contribution are insignificant. To verify this, we have analyzed the intensities of the first and second side lobes on either side of the dual peaks (one can choose any side as they are symmetric in nature) as the number of PR domains increases. From Fig. 3b, we observed that the variation in the intensities of the side lobes is negligible with the increment in the number of PR domains distributed in the PROS QPM structure.

However, the nature of the generated SH peaks depends on the locations of the PR domains in the device. The device profile of the four PR domains PROS device is shown in Fig. 3a. Each bullet point indicates the presence of PR domains at a specific position over the length of the device. Since all the PR domains are equally spaced, the difference between the consecutive PR domains is always constant throughout the length; hence, the graph is a straight line. Therefore, this approach of distributing $N$ number of PR domains at equal intervals in the periodic superlattice structure was limited to the generation of only dual SH peaks regardless of its number [29]. Hence, we employed the idea of PR domains distribution at unequal intervals in the periodic structure; there exists a possibility of generating a desired equal intensity of multiple QPM SHG peaks.

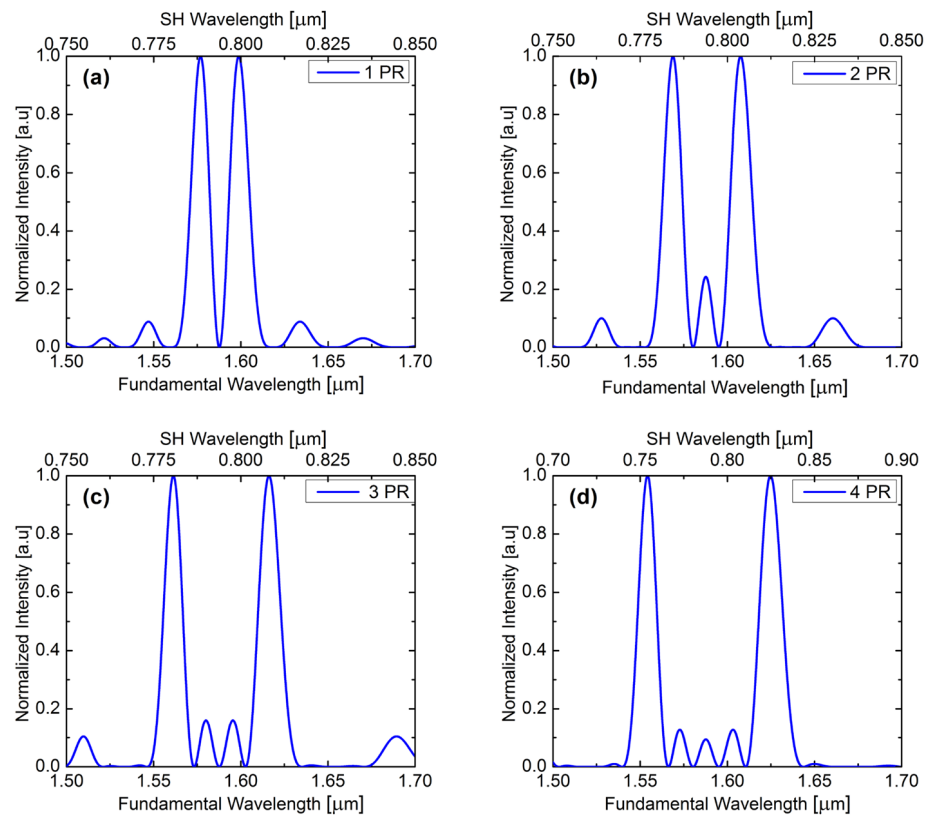

**Fig. 2** Normalized SH spectra of PROS device with $N = 1, 2, 3$ and $4$ PR domains

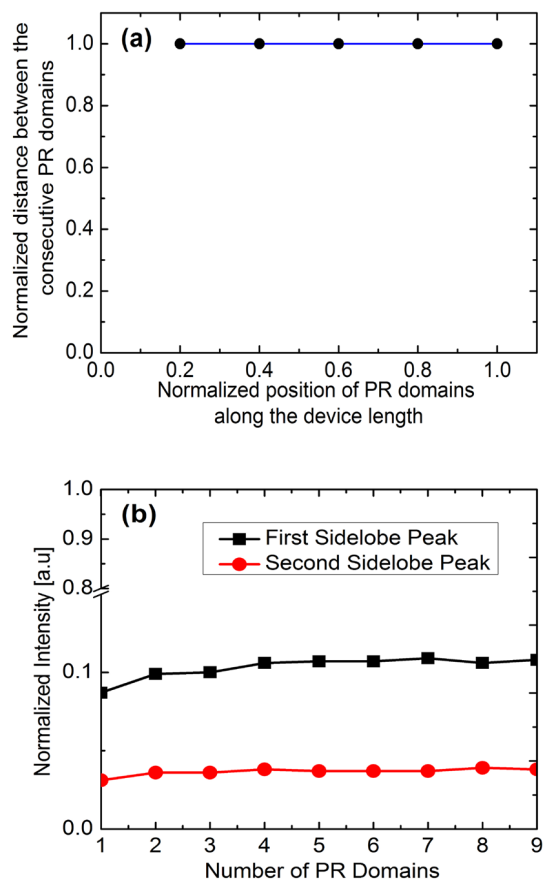

**Fig. 3** **a** Device profile of four PR domains distributed at equal intervals in 10-mm PROS QPM device. **b** Normalized SH intensity of first and second side lobes/bands for different PR domains distributed in the PROS QPM device of 10-mm length

To generate equal-intensity multiple SH peaks, the PR domains must be located at a suitable location in the periodic QPM structure. Five SH peaks (two main and three interleaved peaks) are generated when four PR domains are distributed at equal intervals in the device as shown in Fig. 2d. Instead of distributing the PR domains at the equal intervals, we slightly changed the locations of the first (320th), second (420th), third (520th) and fourth (620th) PR domains. The number in the bracket represents the relative positions of the PR domains distributed in the PROS QPM device. The device profile is shown in Fig. 4a. In this condition, we observed the growth of interleaved peaks intensity with the suppression of side peaks as shown in Fig. 4b. By considering the previous results as a reference, we altered the device profile by changing the locations of the PR domains as shown in Fig. 4c. The new locations of the first, second, third and fourth PR domains are 300th, 460th, 570th and 720th, respectively. This has resulted in the generation of five SH peaks with different intensities as depicted in Fig. 4d. On further fine adjustment of the locations of the first, second, third and fourth PR domains in the device, five SH peaks are generated with almost equal intensities at fundamental wavelengths 1.550 µm, 1.569 µm, 1.588 µm, 1.606 µm and 1.629 µm as shown in Fig. 4f. The final locations of the first, second, third and fourth PR domains are 280th, 440th, 580th and 708th, respectively, and the corresponding device profile is shown in Fig. 4e. Employing this approach, one can generate $N + 1$ multiple QPM peaks by distributing N PR domains at unequal intervals in the PROS QPM device. Besides, the positioning of the PR domains precisely, i.e., identifying suitable location is a tedious job but shows more convenience towards the device fabrication with less duty cycle errors.

Moreover, any variation in the width of the distributed PR domains directly affects the spectral response of the device [30]. Therefore, analyzing the phase variation due to the change of the PR domain width is mandatory as it gives impact on the multiple SHG spectrum. So, we analyzed the effect on the generated SH peaks (five peaks) by changing the width of the PR domains ($N = 4$) in the PROS QPM device. For this analysis, we fixed the locations of four PR domains at 280th, 440th, 580th and 708th in the PROS QPM device.

To get more idea on the variation in frequency conversion spectra due to the change in the width of the PR domain, we have systematically varied the width of the PR domains. In the first case, we vary the width of the PR domains equally (i.e., $w_1 = w_2 = w_3 = w_4 = w$) in the PROS QPM device. The generated SH spectra are shown in Fig. 5. When the width of all the PR domains are equal to $\Lambda/2$, i.e., $w = 10$ µm, a single SH peak was observed due to the zero phase shift, and this is the indication of no PR domains present in the device [31]. The single SH peak is shown in Fig. 5a, and we named it as main peak M. As w increases from $\Lambda/2$ (i.e., $w > 10$ µm), it induces an additional imbalanced phase shift in the PROS resulting to the generation of multiple peaks. So, we observed five unequal peaks for the distribution of four PR domains of each width $w = 3\Lambda/2 = 15$ µm and named as B, A, M, C, and D peaks as shown in Fig. 5b. Similarly, at $w = 7\Lambda/8 = 17.5$ µm, we observed the growth of B, A, M and C peaks as depicted in Fig. 5c. Then, at $w = \Lambda = 20$ µm, the intensities of B, A, M, C and D peaks are equal which is shown in Fig. 5d. In Fig. 5e–g, the peak intensities are gradually decreased due to the change in PR domains widths more than (i.e., $w > 20$ µm). Now, at $w = 3\Lambda/2 = 30$ µm, a single SH peak was generated which is the same as observed in the case of $w = \Lambda/2 = 10$ µm. This indicates the symmetrical nature of the SH responses of the PROS QPM devices.

Maximum intensity was observed at $w = \Lambda/2$ µm or $3\Lambda/2$ µm, leading to the formation of a single peak M

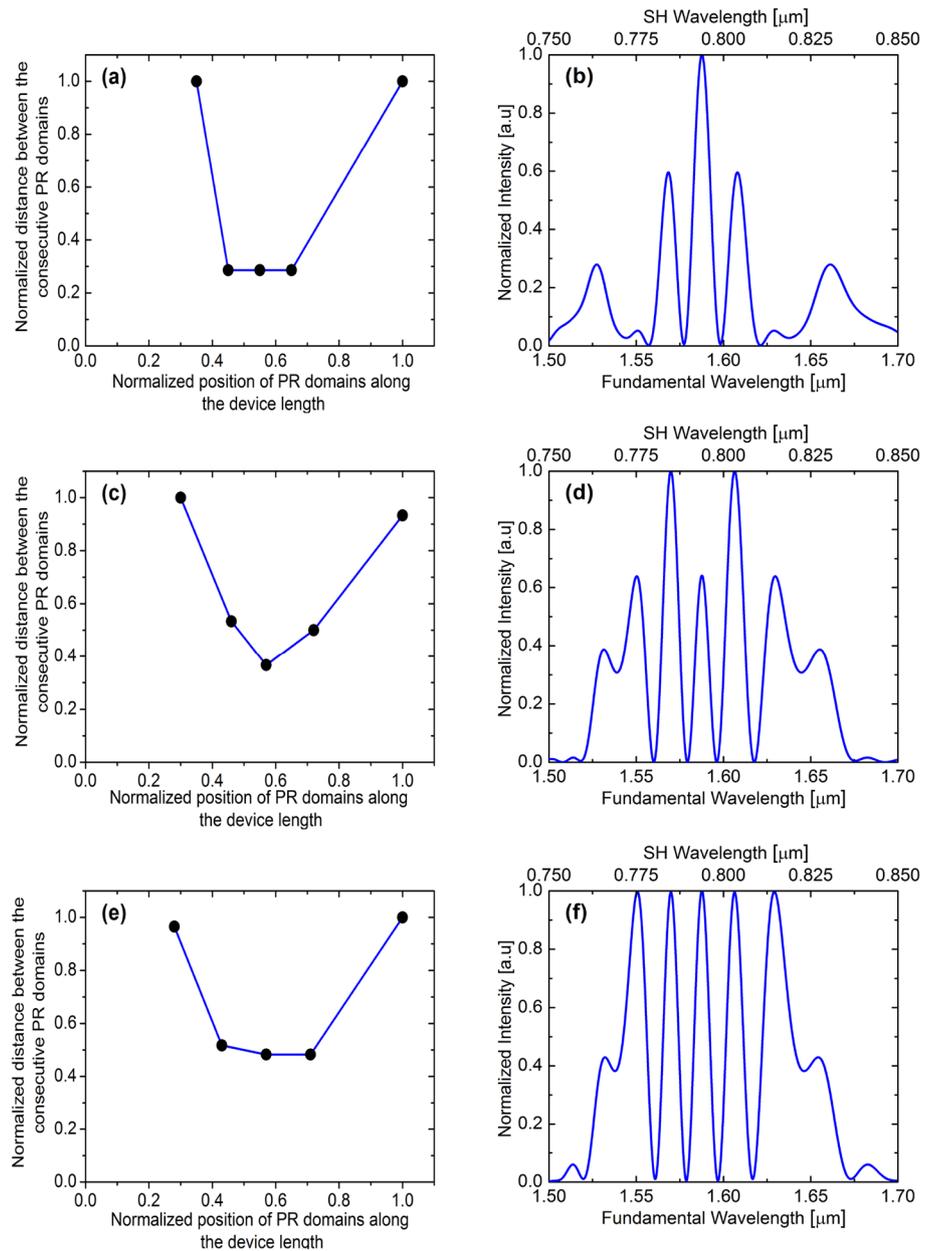

**Fig. 4** The PROS QPM device profiles and the corresponding SH spectra for the formation of five peaks with four PR domains distributed at a different location on the periodic device of 10-mm length

(periodic behavior) by suppressing other peaks B, A, C and D. This analysis was carried out by studying the intensity variation of B, A, C and D peaks with respect to the peak M as shown in Fig. 6. As w increases from $3\Lambda/2$ to $7\Lambda/8$, the intensity of the B, A, C and D are increasing because of the energy distribution from M peak. The formation of five equal intensities peaks was observed at $w = \Lambda$, where the difference of the intensities of all the peaks was nearly zero. On further increasing w more than $\Lambda$, the variation in the intensities of B, A, C and D peaks are exactly opposite to that of the variation observed at w increased from $3\Lambda/4$ to $7\Lambda/8$. This clearly justifies the symmetrical nature of the peaks as depicted in Fig. 6. Nonetheless, it is very obvious that equal intensities of the SHG spectra can be observed only when all the PR domains located are of equal widths same as the grating period of the PROS device. Furthermore, allocation of dissimilar PR domains widths in the PROS device also affected the nature of the generated SHG. Various factors like duty cycle variations, mis-registration of periodicity, photolithographic imperfections and improper poling are inevitable in the electric field poling process that directly suppresses the conversion efficiency of the device [32]. Moreover, the inclusion of PR domain in the periodic structure reduces the conversion efficiency as the energy is equally distributed to other peaks. Its been also used to supress the unwanted parasitic interaction in QPM

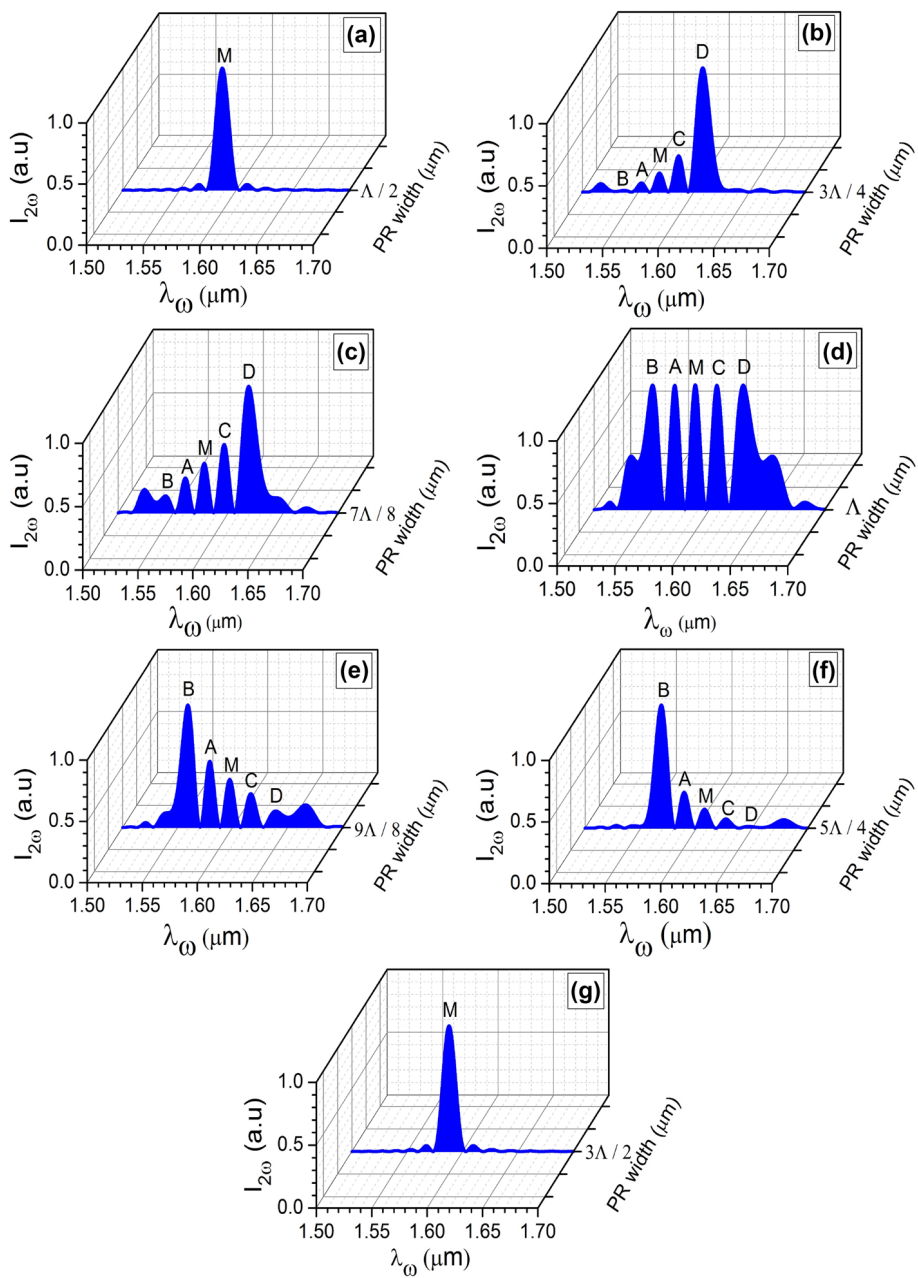

**Fig. 5** SH response generated by PROS QPM device by changing the width of all four PR domains equally at fixed positions in the PROS device of length 10 mm

OPO process [33]. To give an insight into the conversion efficiency of the generated multiple QPM SHG, relative efficiency was calculated by comparing with the perfectly phase-matched or ideal QPM device which can be given by [34]

$$\eta = \frac{I_{\text{multiple QPM}}}{I_{\text{ideal QPM}}} \times 100\%, \qquad (10)$$

where $I_{\text{multiple QPM}}$ and $I_{\text{ideal QPM}}$ are the SH intensities of multiple and ideal QPM devices respectively. Fig. 7 shows the normalized conversion efficiencies of both ideal and five peaks PROS QPM devices of 10-mm length. It was observed that the relative conversion efficiency of five peaks PROS QPM device is $\eta = 18.65\%$. The conversion efficiency can be further improved by choosing the appropriate effective device length. Moreover, the conversion efficiency is also a factor of input pump power and the quality of the QPM device. But, there exists a trade-off between conversion efficiency and the bandwidth of the device [35]. Furthermore, the proposed QPM device possesses the simplest structure which made it comparatively much easier to fabricate with less wastage than the existing available structures. Room temperature electric field poling technique can be used to realize this device [36]. In addition, the number of PR domains N is predefined to generate any desired equal-intensity peaks which are highly flexible and beneficial in the wavelength conversion for future photonic networks.

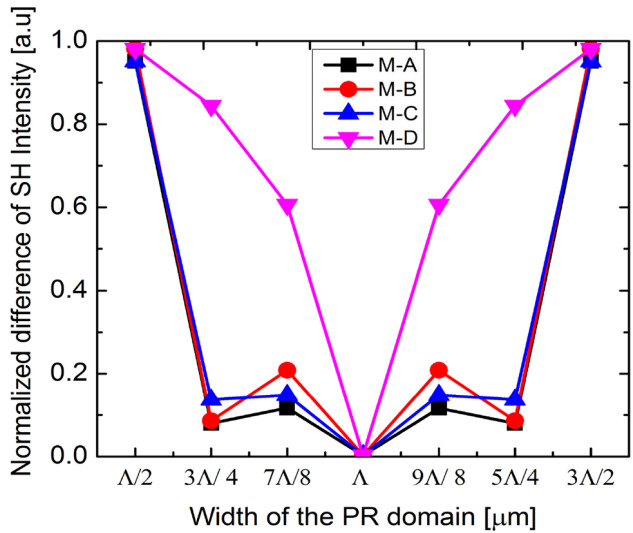

**Fig. 6** Observation of intensity variation of five peak SH spectrum when width of all the distributed four PR domains varied equally

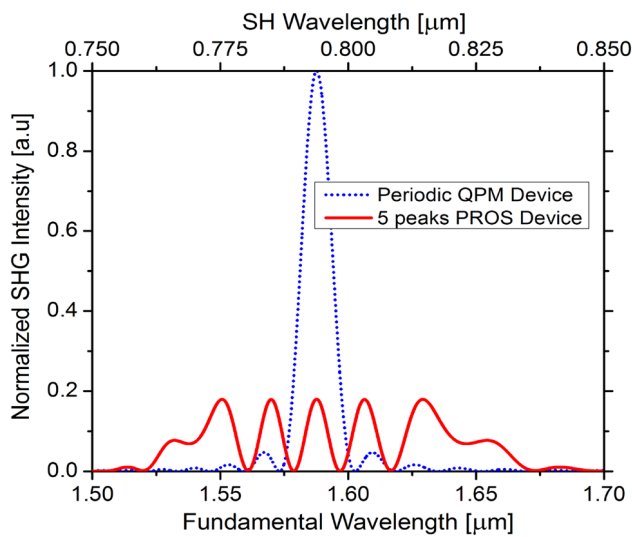

**Fig. 7** Normalized SHG efficiency of ideal (dashed line) and multiple peaks (solid line) QPM devices

## 4 Conclusion

Using the proposed PROS structure, multiple QPM SHG are successfully generated by distributing PR domains in the device. The PR domains are distributed at equal and unequal intervals along the length of the device. The location of the PR domains is very important as it creates an unbalanced phase shift which resulted in the generation of multiple peaks. The distribution of any number of PR domains ($N = 1, 2, 3$) at equal intervals in the device is limited to the generation of equal intensity dual peak SHG. However, low intensity interleaved peaks observed in this scheme when the number of PR domain is greater than one i.e., $N > 1$, contributes to the possibility of generating equal intensity multiple SH peaks. So, four PR domains distributed at unequal intervals in the suitable locations of the PROS QPM device; five equal-intensity multiple SHG peaks are generated for the fundamental wavelengths at 1.550 μm, 1.569 μm, 1.588 μm, 1.606 μm and 1.629 μm in a 10-mm-long device of 20 μm period. The optimal locations of the four PR domains are 280th, 440th, 580th and 708th. The frequency conversion spectra of the device with the variation of the widths of PR domains were analyzed. Any deviation of PR domain width less or greater than the period directly altered the nature and structure of the generated SH peaks but exhibited symmetrical peaks patterns in complementary order.

**Acknowledgements** This work was supported by the Science and Engineering Research Board (SERB), Department of Science & Technology (DST), New Delhi, India under the Grant number CRG/2018/001788.